\documentstyle[pre,aps,preprint,epsf]{revtex}
\begin{document}
\draft
\title{Entropic Elasticity at the Sol-Gel Transition}
\author{Oded Farago and Yacov Kantor}
\address{School of Physics and Astronomy, Raymond and Beverly
  Sackler Faculty of Exact Sciences,\\ Tel Aviv University, Tel
  Aviv 69 978, Israel}
\maketitle
\begin{abstract}
  The sol-gel transition is studied in two purely entropic models
  consisting of hard spheres in continuous three-dimensional space,
  with a fraction $p$ of nearest neighbor spheres tethered by
  inextensible bonds. When all the tethers are present ($p=1$) the two
  systems have connectivities of simple cubic and face-centered cubic
  lattices. For all $p$ above the percolation threshold $p_c$, the
  elasticity has a cubic symmetry characterized by two distinct shear
  moduli. When $p$ approaches $p_c$, both shear moduli decay as
  $(p-p_c)^f$, where $f\simeq 2$ for each type of the connectivity.
  This result is similar to the behavior of the conductivity in random
  resistor networks, and is consistent with many experimental studies
  of gel elasticity. The difference between the shear moduli that
  measures the deviation from isotropy decays as $(p-p_c)^h$, with
  $h\simeq 4$.  
\end{abstract}
Gels are macroscopically large networks, formed when short polymeric
units in a solution (sol) are randomly cross-linked. The transition
from sol to gel is a second order phase transition from fluid to solid
during which the viscosity of the system diverges and shear elasticity
(rigidity) develops. Frequently, the geometry of the gels is modeled
by percolation \cite{stauffer}, when the monomers are represented by
the vertices of some lattice, while the random chemical bonds are
modeled by bonds joining the vertices with probability $p$. The gel
point is identified with the percolation threshold $p_c$, the critical
bond concentration above which a spanning cluster is formed. Close to
$p_c$ quantities like the average cluster size or the gel fraction
have power law dependence on $(p-p_c)$ with exponents independent of
the details of the lattice. Experimental measurements \cite{adam} of
the {\em geometric}\/ features of gels confirm the correspondence with
the percolation model. Therefore, we expect that model systems whose
geometry is described by percolation will produce a correct
description of the {\em physical}\/ properties of gels, such as
elasticity. In this work, we study the elastic properties of a {\em
  purely entropic}\/ percolation model of gels, and compare the
results with experimental measurements of elasticity near the sol-gel
transition, as well as with the predictions of approximate theories.

Close to $p_c$ the shear modulus of a gel follows a power law:
$\mu\sim(p-p_c)^f$. Since the polymeric network forming the gel is
tenuous and floppy, the dominant contribution to its shear modulus is
of entropic origin: Upon distortion of the system, the available phase
space, namely entropy, of the gel decreases, leading to an increase of
the free energy, and to a restoring force. De Gennes suggested that
the exponent $f$ should be equal to the exponent $t$ describing the
conductivity $\Sigma$ of random resistor networks near $p_c$:
$\Sigma\sim(p-p_c)^t$ \cite{degennes}. The equality $f=t$ can be
proved rigorously for a phantom network [without excluded volume (EV)
interactions] of Gaussian springs each having the energy
$E=\frac{1}{2}Kr^2$, where $r$ is the spring length \cite{gaussian}.
It describes the entropic elasticity of most phantom networks since
the latter exhibit an {\em effective}\/ Gaussian behavior on
sufficiently large scales (provided that they are not strongly
stretched) \cite{plischke,phantom}. It is an open question whether the
equality $f=t$ is also valid in the presence of EV interactions. It
does, according to some theories which claim that EV interactions
primarily influence the bulk (compression) modulus rather than the
shear (rigidity) modulus of the system, as if the gel is a phantom
network embedded into a ``pressure producing'' fluid medium
\cite{alexander}. However, a different approach based on scaling
arguments concludes that the elastic moduli of a gel are of the order
of $kT/\xi^d$, where $\xi$ is the percolation correlation length that
diverges as $(p-p_c)^{\nu}$, and $d$ is the dimensionality.
Consequently, the relation $f=d\nu$ is obtained \cite{daoud}.

The experimental values of $f$, measured for different gel systems,
are divided between the above two approaches. In one group of
experiments \cite{firstgroup}, done on materials like gelatin and
silica gels, the measured exponent is close to the conductivity
exponent $t\simeq 2$ in three dimensions \cite{3dconductivity}.
Another group consists of experiments in materials like polyester and
PVC, where the exponent varies from 2.5 to 3.0, and seems to agree
with $f=d\nu\simeq 2.7$ \cite{secondgroup}. The gels formed by the
materials in both groups of experiments are floppy, and the dominance
of the entropic contribution to their elastic properties is fairly
expected. Thus, the division of experimental works into these two
groups is based on the values of the measured exponents rather than on
the nature of the investigated materials. The origin of the
discrepancy between the experimental results is not clear, and we can
only list several possible reasons: In some cases the topology of the
system does not correspond to three-dimensional (3D) percolation model
of gels, but is somewhere between gel (cross-linking of monomers or
short polymeric units) and rubber (cross-linking of a melt of long
polymers) \cite{remark}. Additional reasons are related to
experimental difficulties, such as the imprecise determination of
concentration of cross-links, or the difficulty to extract the static
shear modulus from measurements of the low frequency behavior of the
dynamic complex modulus. A more fundamental reason for the wide range
of experimental results is the energetic contribution to gel
elasticity which mixes with the entropic contribution and influences
the ``effective'' exponent. Energetic {\em bending elasticity}\/ is
characterized by a much larger exponent, $f\simeq 3.8$ \cite{bending}.
Such an exponent is measured only when the entropic contribution to
elasticity is negligible, e.g, in the experiments in sintered metallic
powders \cite{thirdgroup}. When both energetic and entropic
contributions coexist, we expect the elastic behavior near the gel
point to be dominated by the latter, since the critical exponent of
entropic elasticity (according to both approaches to entropic
elasticity) is smaller than that of bending elasticity. However, the
dominance of entropic elasticity near the transition may be limited to
a very narrow regime, in which the shear modulus is small and
difficult to measure.

In the present work we bypass the problem of mixing of the entropic
and elastic contributions and investigate the elastic behavior of
purely entropic systems: We consider a system consisting of hard
spheres, connected by ``tethers'' that have no energy but simply limit
the distance of a connected pair to be smaller than some value $b$. We
apply a new method in which the stress tensor $\sigma_{ij}$ and the
elastic constants $C_{ijkl}$ of such ``hard-spheres-and-tethers''
systems are measured from the probability densities of contact between
spheres and the probability densities of having stretched tethers
\cite{fluctuation}. The stress and elastic constants are the
coefficients of the expansion of the free energy density in the
components of the Lagrangian strain tensor, $\eta_{ij}$:
$f(\{\eta\})=f(\{0\})+\sigma_{ij}\eta_{ij}+{1\over 2}C_{ijkl}\eta_{ij}
\eta_{kl}+\ldots$\,. In this paper we study systems whose elastic
properties posses a cubic symmetry under uniform external pressure
$P$. Such systems have a diagonal stress tensor:
$\sigma_{ij}=-P\delta_{ij}$, and only three {\em different}\/
non-vanishing elastic constants: $C_{11}\equiv
C_{xxxx}=C_{yyyy}=C_{zzzz}\,$; $C_{12}\equiv
C_{xxyy}=C_{yyxx}=C_{yyzz}=\ldots\,$; and $C_{44}\equiv
{1\over2}(C_{xyxy}+C_{xyyx})=\ldots\,$ \cite{wallace}. Cubic systems
have {\em two} shear moduli $\mu_1=C_{44}-P$ and
$\mu_2=\frac{1}{2}(C_{11}-C_{12})-P$.

\begin{figure}[htb]
\epsfysize=18\baselineskip
\centerline{\hbox{
      \epsffile{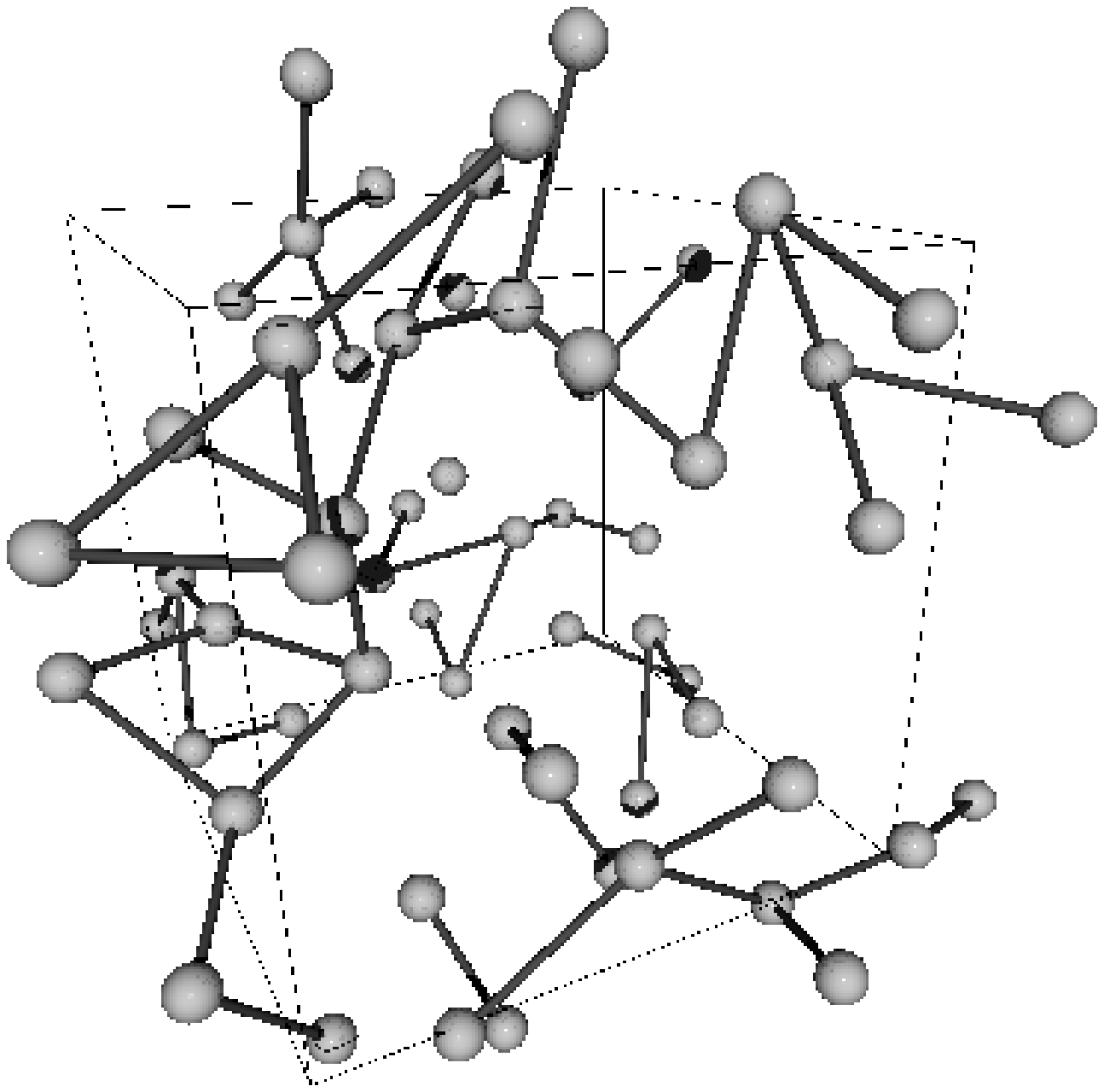}  }}
\caption {\protect A part of an equilibrium configuration of the FCC
  bond percolating system with $p=0.1975$. For clarity the spheres are
  shown as $\frac{1}{3}$ of their actual diameter.}
\label{configuration}
\end{figure}

The topologies of the networks were defined by considering bond
percolation problems on simple cubic (SC) and faced-centered cubic
(FCC) lattices, with a fraction $p$ of bonds present. Each site of the
lattice was occupied by a sphere of diameter $a$, while each present
bond was replaced by a tether of maximal extension $b$, which was
larger than the nearest-neighbor distance $b_0$. Once the topology
(connectivity) was defined, the systems were allowed to move in a {\em
  continuous}\/ 3D space. For both types of topologies we set the
volume fraction of the spheres to be 0.2 and the ratio $b/a\sim 1.6$.
We measured the elastic behavior as a function of $p$. It should be
noted that even in the absence of tethers a dense system solidifies
into FCC solid. In such a case we have a first order phase transition,
where the liquid-solid coexisting phase appears at volume fraction of
spheres 0.494, and the system becomes solid at volume fraction 0.545
\cite{phase}. The topologies of the SC and the FCC systems are quite
different: In the latter the number of nearest-neighbor lattice sites
is larger and, consequently, the percolation threshold is smaller:
$p_c\simeq0.12$ and $p_c\simeq0.249$ for the FCC and SC topologies,
respectively. Thus, highly connected rigid regions are formed more
rapidly (at lower $p$) in FCC networks.  It has been suggested (see,
e.g., Devreux {\em et al.}\/ in Ref.~\cite{firstgroup}) that in real
gels the creation of such rigid blobs tends to enhance the
contribution of energetic bending elasticity and, thus, makes the
entropy-dominated regime near $p_c$ narrower.  Therefore, it seems
interesting to compare the SC and the FCC topologies in a purely
entropic model. In our Monte Carlo (MC) simulations we used box sizes
$L=18b_0$ and $L=12\sqrt{2}b_0$ for the SC and FCC topologies,
respectively, with periodic boundary conditions. Figure
\ref{configuration} depicts a typical equilibrium configuration of the
FCC system. We measured $\sigma_{ij}$ and $C_{ijkl}$ over a broad
range of concentrations above $p_c$. Strictly speaking, the rigidity
threshold $p_r$ is lower than the percolation threshold $p_c$ due to
effects of entanglements \cite{entangelments} (and, perhaps, also due
to additional EV effects). However, the two thresholds are so
extremely close that they are practically indistinguishable in
experiments and numerical studies. Therefore, we treat $p_r$ and $p_c$
as identical. The number of quenched topologies and the length of the
MC run of each individual topology increased as we approached $p_c$.
For systems close to $p_c$ we needed to average the relevant
quantities over 10 different topologies, while far from $p_c$, 2--3
sufficed. Close to $p_c$ the duration of the MC runs is about 500
times larger than the relaxation time $\tau$ of the simulations (see
an approximate expression for $\tau$ in Ref.~\cite{phantom}). During
each MC run the systems were sampled several million times.

\begin{figure}[htb]
\epsfysize=16\baselineskip
\centerline{\hbox{
      \epsffile{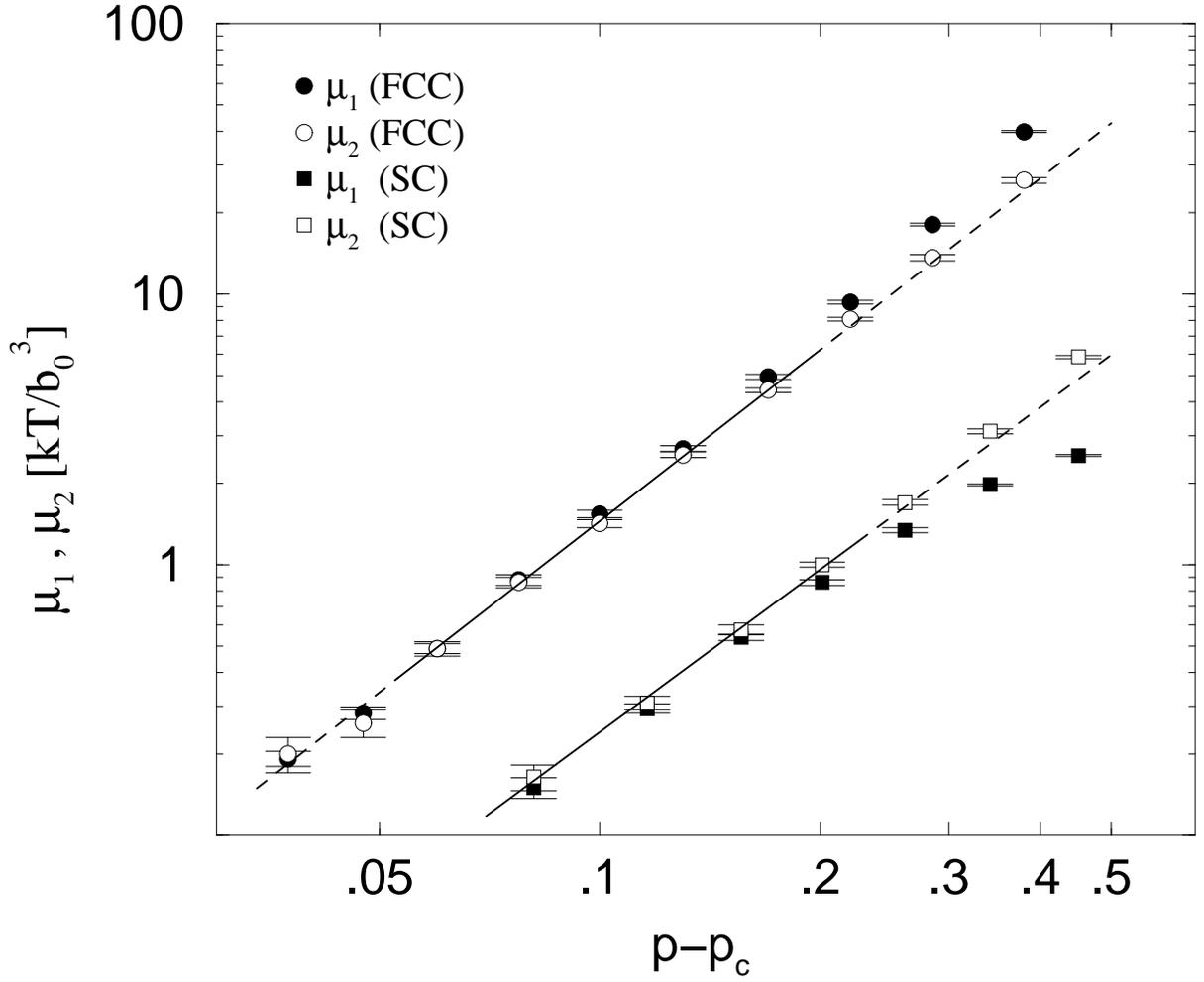}  }}
\caption {\protect Logarithmic plot of the shear moduli $\mu_1$ (solid
  symbols) and $\mu_2$ (open symbols) as a function of $(p-p_c)$, for
  FCC (circles) and SC (squares) bond percolating systems. For each
  topology, its percolation threshold is used ($p_c\simeq0.12$
  for FCC, and, $p_c\simeq0.249$ for SC). For both systems the
  volume fraction is 0.2 and $b/a\sim 1.6$.}
\label{shear}
\end{figure}

\begin{figure}[htb]
\epsfysize=16\baselineskip
\centerline{\hbox{
      \epsffile{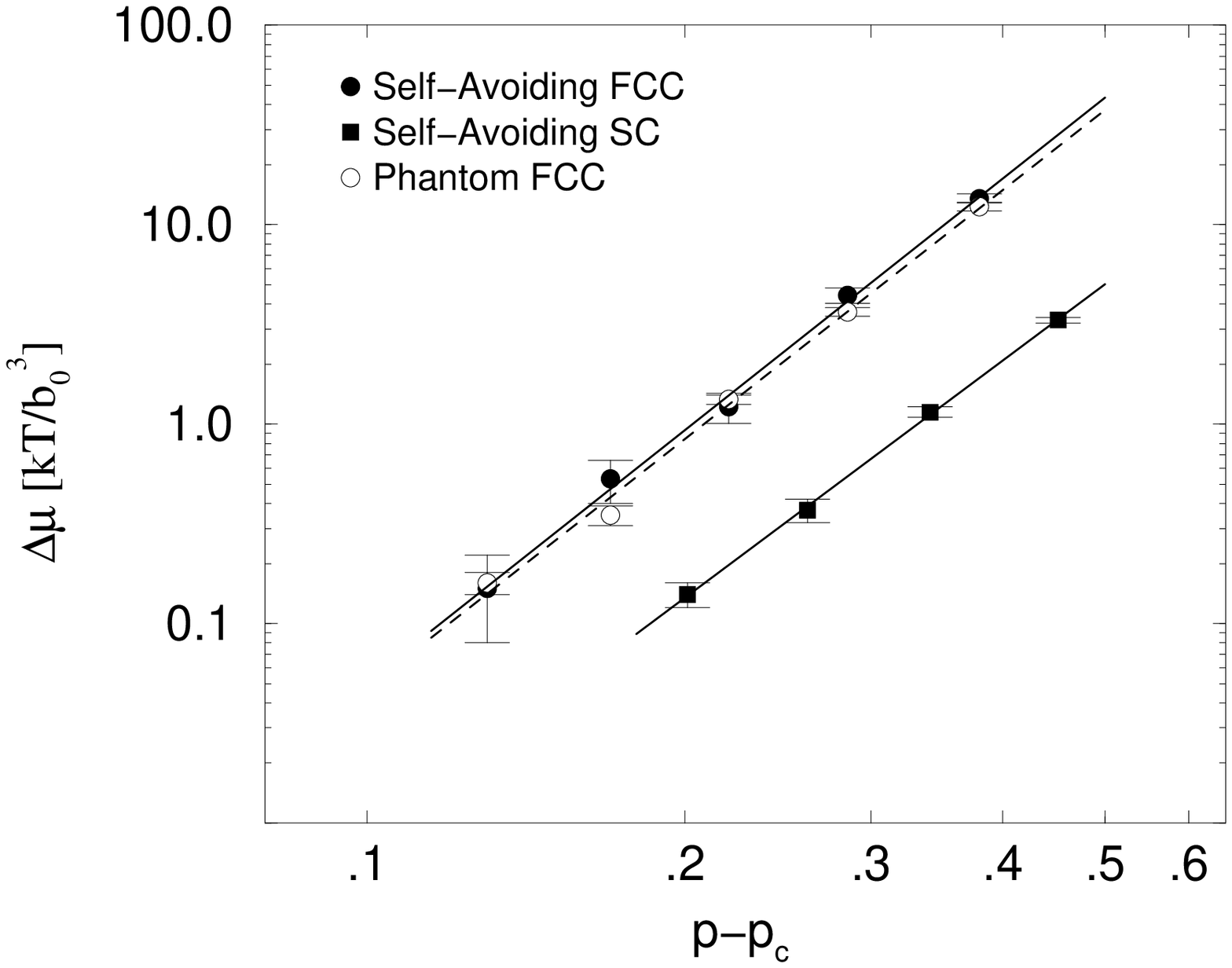}  }}
\vspace{1cm}
\caption {\protect Logarithmic plot of the difference between the
  shear moduli $\Delta\mu$ as a function of $(p-p_c)$,
  for SA FCC (solid circles), SA SC (squares) and phantom FCC (open
  circles) bond percolating systems.}
\label{iso}
\end{figure}

The networks studied in this work posses a cubic symmetry since their
topologies are defined on cubic lattices. Therefore, their elastic
behavior is described by two distinct shear moduli rather than one, as
in isotropic systems. This property does not exist in experiments
where the networks are isotropic because of randomness. Figure
\ref{shear} depicts the two shear moduli, $\mu_1$ and $\mu_2$, as a
function of $(p-p_c)$ for the SC and FCC systems. For each type of
connectivity, its own $p_c$ is used. The error bars appearing in
Figs.~\ref{shear} and \ref{iso} correspond to one standard deviation
of the averaged quantities. For both systems close to $p_c$, $\mu_1$
and $\mu_2$ are practically indistinguishable, suggesting that the
systems become isotropic. The shear moduli can be approximated by the
power laws $\mu_1\simeq\mu_2\sim(p-p_c)^f$ with $f=2.0\pm0.1$ for the
SC system, and $f=2.1\pm0.1$ for the FCC system. Within numerical
uncertainty both values are similar and consistent with recent
estimates of the conductivity exponent in 3D, $t\simeq2.0$
\cite{3dconductivity}. We already found that $f\simeq t\simeq 1.3$ in
two dimensions \cite{self2d} and, therefore, we expect that $f\simeq
t$ at any dimension. As we have mentioned earlier in the text, the
equality $f=t$ is expected for a phantom network ($a=0$) and can be
explained by the Gaussian nature of its elastic response close to
$p_c$. Our results indicate that a similar picture may apply to
systems with EV interactions.

In Fig.~\ref{shear} we observe that the values of $\mu_1$ and $\mu_2$
gradually deviate from each other far from $p_c$ because at large $p$
the systems ``remember'' the lower (cubic) symmetry of their
connectivities. For the FCC connectivity $\mu_1>\mu_2$, while for the
SC case $\mu_2>\mu_1$. (The definitions of the shear moduli $\mu_1$
and $\mu_2$ depend on the orientation of the axes of the reference
system, which in our study were taken along the edges of the
conventional cubic unit cell.)  Figure \ref{iso} shows that the
difference $\Delta\mu\equiv|\mu_1-\mu_2|$, follows, in both cases,
quite similar power laws $\Delta\mu\sim(p-p_c)^h$ with $h=3.95\pm
0.15$ for the SC case and $h=4.15\pm0.15$ for the FCC case. Because of
the similarities of the values of $h$ in SC and FCC systems, it is
reasonable to assume that $h$ is a new universal critical exponent
which characterizes deviation from isotropic elastic behavior. While
the power law dependence of $\Delta\mu$ is not surprising due to the
self-similar nature of the large percolation clusters, we could only
support this assumption by numerical data of limited accuracy. We
verified the validity of the power law dependence on $(p-p_c)$ by
attempting (unsuccessfully) to fit the data to other functional forms.

We already saw that the exponent $f$ (describing the leading critical
elastic behavior) is very similar for self avoiding (SA) and phantom
percolating systems. Therefore, it is interesting to check whether
this similarity applies to the exponent $h$, as well. For this purpose
we measured $\Delta\mu$ for a phantom FCC bond percolating network
with the same values of $b$ and $b_0$, but with $a=0$ \cite{remark2}.
The results of these simulations are also plotted in Fig.~\ref{iso},
revealing a power law with $h=4.15\pm0.15$, as in the SA FCC case. The
phantom Gaussian model which predicts that $f=t$, cannot be used to
predict the value of $h$ since it gives $\Delta\mu \equiv 0$ at any
bond concentration $p$ \cite{gaussian,gaussian2}. Hence, $\Delta\mu$
represents deviation from a purely Gaussian behavior which originates
in the non-Gaussian form of the tether potential and (in the SA case)
EV interactions. Our results for the exponent $h$ imply that the
similarity between the critical elasticity of phantom and SA
percolating systems may not be restricted to the leading Gaussian
behavior.

In conclusion, we have studied the entropic elasticity of 3D purely
entropic percolating systems. Our study shows that the the critical
behaviors of the shear moduli of phantom and SA percolation systems
are characterized by similar critical exponents which are very close
to the conductivity exponent. (For phantom systems the elasticity
exponent actually coincides with the conductivity exponent.) This
result agrees with many experimental studies of gel elasticity. It
corresponds to heuristic theories which assume that the finite
clusters (1) do not contribute directly to the shear modulus (i.e.,
behave like a fluid medium) and (2) effectively screen out EV
interactions in the elastic network. Further support to this
theoretical description is given by our result for $\Delta\mu$ which
is also described by similar exponents in phantom and SA systems. The
exponent $h$ that characterizes the decay of $\Delta\mu$ seems to be
universal, namely independent of the lattice on which the geometry of
the system is defined, but this point should be established more
carefully by studying other lattice connectivities, and by measuring
$h$ for two dimensional (phantom and SA) percolation systems.

We thank M.~Kardar for numerous discussions of the problem. This work
was supported by the Israel Science Foundation through Grant No.
177/99. The numerical simulation were performed on SGI Origin2000
supercomputer and Beowulf cluster at the High Performance Computing
Unit at the Inter University Computation Center of Israel. We thank
G.~Koren for his support in conducting the numerical work.




\begin{thebibliography}{99}
\bibitem{remark2} In the
  phantom case we simulated only the backbone of the spanning cluster
  since the finite clusters and the dangling ends of the spanning
  cluster do not influence the shear moduli. See Refs.~{\protect
    \cite{gaussian,phantom}} where we discuss the elastic behavior of
  phantom networks in more details.
\end{thebibliography}
\end{document}